\documentclass[a4paper,12pt]{article}
\usepackage{amssymb,amsmath}
\usepackage{fancyhdr,appendix,verbatim}
\usepackage{parskip}
\long\def\del#1\enddel{}
\newcommand{\sect}[1]{\setcounter{equation}{0}\section{#1}}

\numberwithin{equation}{section}

\def\V{{\mathrm V}}

\newcommand{\am}{\mbox{\bf am}}
\newcommand{\sn}{\mbox{\bf sn}}

\def\axs{{\rm AdS}_5\times S^5}
\newcommand{\eq}[1]{\begin{equation} #1 \end{equation}}
\newcommand{\al}[1]{\begin{align} #1 \end{align}}

\begin{document}
\begin{titlepage}
\title{Three-point correlation functions from pulsating strings in AdS$_5\times S^5$}
\author{D.~Arnaudov${}^{a}$\thanks{\texttt{dlarnaudov@phys.uni-sofia.bg}}\ \ and R.~C.~Rashkov${}^{b,a}$\thanks{\texttt{rash@hep.itp.tuwien.ac.at}}
\ \\ \ \\
${}^a$ Department of Physics, Sofia University,\\
5 J. Bourchier Blvd, 1164 Sofia, Bulgaria
\ \\ \ \\
${}^b$ Institute for Theoretical Physics,\\
Vienna University of Technology,\\
Wiedner Hauptstr. 8-10, 1040 Vienna, Austria
}
\date{}
\end{titlepage}

\maketitle

\begin{abstract}
One of the most important problems in any conformal field theory is the calculation of three-point functions of primary operators. In this paper we provide explicit examples of correlators with two scalar operators in ${\cal N}=4$ super-Yang--Mills theory at large $N$, corresponding to pulsating semiclassical strings in AdS$_5\times S^5$, and an operator with small quantum numbers at strong coupling.
\end{abstract}

\sect{Introduction}

An extremely active area of research in theoretical high-energy physics in recent years has been the correspondence between gauge and string theories. Following the impressive conjecture made by Maldacena \cite{Maldacena} that type IIB string theory on $\axs$ is dual to ${\cal N}=4$ super-Yang--Mills theory with a large number of colors, an explicit realization of the AdS/CFT correspondence was provided in \cite{GKP}. Many convincing results have been achieved thereafter, paving the way for the subject to become an indispensable tool in probing such diverse areas as the dynamics of quark-gluon plasma and high-temperature superconductivity.

A key feature of the duality is the connection between planar correlation functions of conformal primary operators in the gauge theory and correlators of corresponding vertex operators of closed strings with $S^2$ worldsheet topology. Recently, some progress was accomplished in the study of three- and four-point functions with two and three ``heavy'' vertex operators with large quantum numbers at strong coupling. The remaining operators were chosen to be various ``light'' states (with quantum numbers and dimensions of order one). It was shown that the large $\sqrt{\lambda}$ behavior of such correlators is fixed by a semiclassical string trajectory governed by the heavy operator insertions, and with sources provided by the vertex operators of light states.

Initially this approach was utilized in the calculation of two-point functions of heavy operators in \cite{Polyakov:2002}--\cite{Buchbinder:2010vw}. More recently the above procedure was extended to certain three-point correlators in \cite{Janik:2010gc,Zarembo:2010,Costa:2010}. A method based on heavy vertex operators was proposed in \cite{Roiban:2010}. Further developments in the computation of correlators with two string states are presented in~\cite{Hernandez:2010}. The main goal of these investigations is elucidation of the structure of three-point functions of three semiclassical operators \cite{Klose:2011}.

Recently the authors of \cite{Bajnok:2014} noticed that the precise formulation of such correlators should involve string energy eigenstates, which necessitates a slight modification of previous methods. Namely, one should average over all string solutions with a given energy. Although this alteration does not invalidate the results for the correlation functions obtained so far, it turns out that in the case of pulsating strings \cite{Minahan:2002rc,Engquist:2003rn} we need to apply strictly the procedure, described in \cite{Bajnok:2014}, in order to get the correct answer. In the present paper we consider the three-point correlation function of two heavy operators, corresponding to a pulsating string solution in $S^3\subset S^5$ \cite{Kruczenski:2004}, and one BPS (dilaton or chiral primary) operator. We provide some limiting cases and recover known results.

The paper is organized in the following way. In Section \ref{sec2} we present a brief review of the procedure for calculating semiclassically two- and three-point correlation functions via vertex operators. In Section \ref{sec3} we proceed with the derivation of three-point correlators for a particular pulsating string solution, taking either the dilaton or the chiral primary operator~(CPO) as the light operator. We study a number of limiting cases of the correlation functions. We conclude with a short discussion on the results.

\sect{Correlation functions with two heavy operators}\label{sec2}

We commence with a review of the method for obtaining two-point correlation functions. Their computation in the leading semiclassical approximation is closely related to utilizing an adequate classical string solution \cite{Tseytlin:2003}--\cite{Janik:2010gc}. If $V_{H1}(\xi_1)$ and $V_{H2}(\xi_2)$ are the two heavy vertex operators, which are inserted at the $\xi_1$ and $\xi_2$ points on the string worldsheet, the corresponding two-point correlator in the limit of large 't Hooft coupling is obtained from the stationary point of the action
\eq{
\langle V_{H1}(\xi_1)V_{H2}(\xi_2)\rangle\sim e^{-I},
}
where $I$ is the action of the $\axs$ string sigma model in the usual embedding coordinates
\al{
&I=\frac{\sqrt{\lambda}}{4\pi}\int d^2\xi\ \Big(\partial Y_M\bar{\partial}Y^M+\partial X_k\bar{\partial}X_k+{\rm fermions}\Big)\,,\\
&Y_MY^M=Y_0^2+Y_1^2+Y_2^2+Y_3^2+Y_4^2-Y_5^2=-R^2,\quad\ X_kX_k=1\,.\nonumber
}
Throughout the paper we apply conformal gauge and use a worldsheet with Euclidean signature. Correspondingly, the two-dimensional derivatives are $\partial=\partial_1+i\partial_2,\,\bar{\partial}=\partial_1-i\partial_2$. We also work with the Euclidean continuation of AdS$_5$. The embedding, global and Poincar\'e coordinates of AdS$_5$ assume the following form
\al{
&Y_5+iY_0=R\,\cosh\rho\ e^{it},\quad
Y_1+iY_2=R\,\sinh\rho\,\cos\Theta\,e^{i\phi_1},\quad
Y_3+iY_4=R\,\sinh\rho\,\sin\Theta\,e^{i\phi_2},\nonumber\\
&Y_m=\frac{R\,x_m}{z}\,,\qquad
Y_4=\frac{1}{2z}(-R^2+z^2+x^mx_m)\,,\qquad
Y_5=\frac{1}{2z}(R^2+z^2+x^m x_m)\,,
\label{poincare}
}
where $x^mx_m=x_0^2+x_ix_i\ (m=0,1,2,3;\ i=1,2,3)$.

The stationary solution satisfies the string equations of motion with singular sources given by $V_{H1}(\xi_1)$ and $V_{H2}(\xi_2)$. Utilizing the conformal symmetry of the theory, we are able to map the $\xi$-plane worldsheet to a Euclidean cylinder with $(\tau,\sigma)$ coordinates
\eq{
e^{\tau+i\sigma}=\frac{\xi-\xi_2}{\xi-\xi_1}\,.
\label{confmap}
}
Under this Schwarz--Christoffel mapping the singular solution on the $\xi$-plane goes to a smooth solution on the cylinder \cite{Tseytlin:2003,Buchbinder:2010,Buchbinder:2010vw} with $t=\kappa\tau$, where $\kappa$ is a constant parameter proportional to the string energy. The quantum numbers of the latter solution coincide with the quantum numbers of the heavy vertex operators, guaranteeing that there is no loss of information.

The considerations above can also be applied to a physical integrated vertex operator dependent on a point ${\rm x}$ on the boundary of the Poincar\'e patch of AdS$_5$ \cite{Polyakov:2002,Tseytlin:2003}
\eq{
{\rm V}_H({\rm x})=\int d^2\xi\ V_H(\xi;{\rm x})\,,\qquad V_H(\xi;{\rm x})\equiv V_H(z(\xi),x(\xi)-{\rm x},X_k(\xi))\,.
}
Again the semiclassical two-point correlation function $\langle\V_{H1}({\rm x}_1)\V_{H2}({\rm x}_2)\rangle$ is determined by the classical action evaluated on the stationary point solution. Applying the conformal mapping \eqref{confmap}, we obtain the corresponding smooth spinning string solution in terms of Poincar\'e coordinates, with the boundary conditions\footnote{We refer to \cite{Buchbinder:2010vw} for details.}
\eq{
\tau\rightarrow-\infty\ \ \Longrightarrow\ \ z\rightarrow0\,,\ \ x\rightarrow{\rm x}_1\,,\qquad\tau\rightarrow+\infty\ \ \Longrightarrow\ \ z\rightarrow0\,,\ \ x\rightarrow{\rm x}_2\,.
\label{boundcond}
}

In a similar fashion we can calculate three-point correlation functions with two heavy and one light operators \cite{Zarembo:2010,Roiban:2010}
\al{
G_3({\rm x}_1,{\rm x}_2,{\rm x}_3)&=\langle\V_{H1}({\rm x}_1)\V_{H2}({\rm x}_2)\V_L({\rm x}_3)\rangle\\
&=\int{\cal D}\mathbb{X}^\mathbb{M}\ e^{-I}\int d^2\xi_1d^2\xi_2d^2\xi_3\ V_{H1}(\xi_1;{\rm x}_1)V_{H2}(\xi_2;{\rm x}_2)V_L(\xi_3;{\rm x}_3)\,,\nonumber
}
where $\int{\cal D}\mathbb{X}^\mathbb{M}$ is the integral over $(Y_M,X_k)$. We note that the contribution of the light operator in the stationary point equations can be neglected, so that one can use the same classical string solution as in the case of the two-point function of two heavy operators. In this way we obtain \cite{Roiban:2010}
\eq{
\frac{G_3({\rm x}_1,{\rm x}_2,{\rm x}_3)}{G_2({\rm x}_1,{\rm x}_2)}=\int d^2\xi\ V_L(z(\xi),x(\xi)-{\rm x}_3,X_k(\xi))\,,
\label{strconstxi}
}
where $(z(\xi),x(\xi),X_k(\xi))$ denote the respective string solution with the same quantum numbers as the heavy vertex operators, and with the boundary conditions in \eqref{boundcond} mapped to the $\xi$-plane by the Schwarz--Christoffel mapping \eqref{confmap}. Using the two-dimensional conformal invariance, we can also provide \eqref{strconstxi} in terms of the cylinder ($\int d^2\sigma=\int^\infty_{-\infty}d\tau\int^{2\pi}_0d\sigma$)
\al{
\label{strconstst}
&\frac{G_3({\rm x}_1,{\rm x}_2,{\rm x}_3)}{G_2({\rm x}_1,{\rm x}_2)}=\\
&=\lim_{T\rightarrow\infty}\frac{1}{T}\int_{-T/2}^{T/2}d\tau_0\int d^2\sigma\ V_L(z(\tau-\tau_0,\sigma),x(\tau-\tau_0,\sigma)-{\rm x}_3,X_k(\tau,\sigma))\,e^{-(\Delta_2-\Delta_1)\kappa\tau_0},
\nonumber
}
where, as was detailed in \cite{Bajnok:2014}, we have averaged over all solutions in AdS (parameterized by different values of $\tau_0$) with the same energy in order to obtain the needed energy eigenstate. We have denoted the conformal dimension of $V_{H1}$ with $\Delta_1$ and that of $V_{H2}$ with $\Delta_2$.

The global conformal $\textrm{SO}(2,4)$ symmetry fixes the spacetime dependence of two- and three-point functions\footnote{We assume that $\V_{H2}=\V^*_{H1}$, which is valid for the correlation functions we are interested in.}
\al{\label{2point}
G_2({\rm x}_1,{\rm x}_2)&=\frac{C_{12}\ \delta_{\Delta_1\!,\Delta_2}}{{\rm x}_{12}^{\Delta_1+\Delta_2}}\,,\qquad{\rm x}_{ij}\equiv|{\rm x}_i-{\rm x}_j|\,,\\
G_3({\rm x}_1,{\rm x}_2,{\rm x}_3)&=\frac{C_{123}}{{\rm x}_{12}^{\Delta_1+\Delta_2-\Delta_3}{\rm x}_{13}^{\Delta_1+\Delta_3-\Delta_2}
{\rm x}_{23}^{\Delta_2+\Delta_3-\Delta_1}}\,,
\label{3point}
}
where $\Delta_i$ are the dimensions of corresponding operators. Choosing properly ${\rm x}_i$, we can suppress the dependence on ${\rm x}_{ij}$ in \eqref{strconstst}, and apply the prescription given in \eqref{strconstst} to compute the structure constant $C_{123}$~\cite{Zarembo:2010,Roiban:2010}. Having in mind that $\Delta_1\approx\Delta_2$ and setting $C_{12}=1$ in \eqref{2point}, we determine that
\eq{
\frac{G_3({\rm x}_1,{\rm x}_2,{\rm x}_3=0)}{G_2({\rm x}_1,{\rm x}_2)}=C_{123}\left(\frac{{\rm x}_{12}}{|{\rm x}_1||{\rm x}_2|}\right)^{\Delta_3}\!.
\label{norm3point}
}
For further details we refer the interested reader to \cite{Buchbinder:2010,Buchbinder:2010vw,Roiban:2010,Bajnok:2014}.

\sect{Three-point correlators from pulsating strings in $\mathbb{R}\times S^3$}\label{sec3}

In the present Section we use the approach outlined above for the calculation of specific three-point correlators. Without loss of generality we can fix ${\rm x}_1=(-1,0,0,0)$ and ${\rm x}_2=(1,0,0,0)$, from which follows that $R=1$. We consider a particular pulsating string in $\mathbb{R}\times S^3\subset\axs$ \cite{Kruczenski:2004} as the string solution that describes the semiclassical trajectory. Using that the string energy is $E=\sqrt{\lambda}\kappa$ and the spin is $J=\sqrt{\lambda}{\cal J}$, the solution is defined as
\al{\label{sol}
t&=\kappa\tau\,,\ \ \rho=0\,,\ \ \cos\theta(\tau)=a_-\sn\!\left(ima_+\tau,\frac{a_-}{a_+}\right),\\ \nonumber
\varphi_1(\tau)&=-\frac{\cal J}{ma_+}\Pi\!\left[\am\!\left(ima_+\tau,\frac{a_-}{a_+}\right),a_-^2,\frac{a_-}{a_+}\right],\ \ \varphi_2=m\sigma\,,\\
\kappa^2&=-\dot{\theta}^2+\frac{{\cal J}^2}{\sin^2\theta}+m^2\cos^2\theta\,,\quad a_{\pm}^2=\frac{\kappa^2+m^2\pm\sqrt{(\kappa^2-m^2)^2+4m^2{\cal J}^2}}{2m^2}\,,
\nonumber
}
where we have assumed the notation of \cite{Gradshteyn} for Jacobi elliptic functions, and $(\theta,\varphi_1,\varphi_2)$ parameterize $S^3\subset S^5$ with metric
\eq{
ds_{S^3}^2=d\theta^2+\sin^2\theta\,d\varphi_1^2+\cos^2\theta\,d\varphi_2^2\,.
}
In Poincar\'e coordinates \eqref{poincare} the AdS part of the solution is
\eq{
z=\frac{1}{\cosh[\kappa(\tau-\tau_0)]}\,,\quad x_0=\tanh[\kappa(\tau-\tau_0)]\,,\quad x_i=0\,,
}
where we have left the integration constant $\tau_0$ unfixed, because we will need to average our expressions over it. It can be shown that the above solution possesses the right asymptotic behavior, namely, $\lim_{\tau\rightarrow\pm\infty}z=0$ and $\lim_{\tau\rightarrow\pm\infty}x_0=\pm1$. Note that by taking ${\cal J}=0$ we would get the original solution for pulsating strings in $\mathbb{R}\times S^2$ \cite{Minahan:2002rc}.

We will proceed with the study of the corresponding three-point correlation functions with two heavy and one light operators. We will examine two choices for the light operator~-- dilaton or superconformal primary scalar (chiral primary operator).

\subsection{Dilaton as light operator}
It is known that the ten-dimensional dilaton field is decoupled from the metric in the Einstein frame \cite{Kim:1985}. Consequently, it is described by a free massless ten-dimensional Laplace equation in $\axs$. The respective string vertex operator is proportional to the worldsheet Lagrangian ($j\geq0$ is the $S^5$ momentum of the dilaton)
\al{\label{dilvertex}
V_L({\rm x}=0)&=V^{(\rm dil)}_j(0)=\hat{c}_{\Delta}K_{\Delta}\,{\rm X}^j\big[(\partial x_m\bar{\partial}x^m+\partial z\bar{\partial}z)/z^2+
\partial X_k\bar{\partial}X_k+{\rm fermions}\big]\,,\\
K_{\Delta}&\equiv\left(\frac{z}{z^2+x^mx_m}\right)^{\Delta}\!,\qquad{\rm X}\equiv X_1+iX_2=e^{i\varphi_1},\nonumber
}
where $\hat{c}_\Delta$ is a constant determined by the normalization of the dilaton. The conformal dimension of the dilaton is $\Delta=4+j$ to the leading order in the large 't Hooft coupling expansion. The corresponding operator in the dual gauge theory is proportional to ${\rm tr}(F^2_{mn}Z^j+\ldots)$. For $j=0$ it is given by the SYM Lagrangian.

From \eqref{strconstst}, \eqref{norm3point} and \eqref{dilvertex} we obtain that
\al{
\label{dilstrconst}
C_{123}&=4c_{\Delta}\lim_{T\rightarrow\infty}\frac{1}{T}\int_{-T/2}^{T/2}d\tau_0\int^{\infty}_{-\infty}d\tau
\int^{\pi/2}_0d\sigma\,K_{\Delta}\,U\,e^{-(\Delta_2-\Delta_1)\kappa\tau_0}\\
U&={\rm X}^j\big[(\partial x_m\bar{\partial}x^m+\partial z\bar{\partial}z)/z^2+\partial X_k\bar{\partial}X_k\big],\qquad c_{\Delta}=2^{-\Delta}\hat{c}_{\Delta}\,.
}
The authors of \cite{Roiban:2010} calculated the normalization constant of the dilaton $\hat{c}_\Delta$ as
\eq{
\hat{c}_\Delta=\hat{c}_{4+j}=\frac{\sqrt{\lambda}}{8\pi N}\sqrt{(j+1)(j+2)(j+3)}\,.
}
Evaluating $U$ on the pulsating string solution \eqref{sol}, we get
\eq{
U=(\kappa^2+\dot{\theta}^2-\frac{{\cal J}^2}{\sin^2\theta}+m^2\cos^2\theta)\,e^{ij\varphi_1}=2m^2\cos^2\theta\,e^{ij\varphi_1},
}
so that the expression in \eqref{dilstrconst} takes the following form
\al{\nonumber
C_{123}&=8m^2c_{\Delta}\lim_{T\rightarrow\infty}\frac{1}{T}\int_{-T/2}^{T/2}d\tau_0\int^{\infty}_{-\infty}d\tau
\int^{\pi/2}_0d\sigma\,\frac{\cos^2\theta\,e^{ij\varphi_1-j{\cal J}\tau_0}}{\cosh^{4+j}[\kappa(\tau-\tau_0)]}\\
&=4\pi m^2c_{\Delta}\int^{\infty}_{-\infty}d\tau'\,\frac{e^{-j{\cal J}\tau'}}{\cosh^{4+j}(\kappa\tau')} \lim_{T\rightarrow\infty}\frac{1}{T}\int_{-T/2}^{T/2}d\tau\,\cos^2\theta\,e^{ij\varphi_1-j{\cal J}\tau},
\label{dilstrconst'}
}
where in the integral over $\tau_0$ we have changed the integration variable to $\tau'=\tau_0-\tau$. The first integral in the second line could be computed in terms of hypergeometric functions. The second integral, however, is difficult to calculate analytically due to the presence of an elliptic integral of the third kind in the exponent. Therefore, we will study the structure constant for particular values of the parameters. First, we note that when $m=0$ we get the three-point function with light operator corresponding to a point-like string. In this case, as has been explained in \cite{Kruczenski:2004}, the equation of motion for $\theta$ leads to $\theta=\pi/2$. Thus, it follows that $\kappa={\cal J}$, which means that $E=J$ as expected for a BPS solution. It can be easily seen that if we set $m=0$ in \eqref{dilstrconst'}, we will indeed get a vanishing structure constant. Next, let us concentrate on the most significant case of $j=0$. It can be obtained that
\al{\nonumber
C_{123}&=\frac{16\pi}{3}\frac{m^2}{\kappa}c_{\Delta}\lim_{T\rightarrow\infty}\frac{1}{T}\int_{-T/2}^{T/2}d\tau\,\cos^2\theta\\
&=\frac{16\pi}{3}\frac{m^2a_-^2}{\kappa}c_{\Delta}\lim_{T\rightarrow\infty}\frac{1}{T}\int_{-T/2}^{T/2}d\tau\,\sn^2\!\left(ima_+\tau,\frac{a_-}{a_+}\right).
}
The resulting integral is divergent. In order to obtain a finite result, we analytically continue $m\rightarrow-im$. We will reverse this operation in the end. We get for the integral
\al{\nonumber
C_{123}&=-\frac{16\pi}{3}\frac{m^2a_-^2}{\kappa}c_{\Delta}\lim_{T\rightarrow\infty}\frac{1}{T}\int_{-T/2}^{T/2}d\tau\,\sn^2\!\left(ma_+\tau,\frac{a_-}{a_+}\right)\\
&=-\frac{16\pi}{3}\frac{m^2a_-^2}{\kappa}c_{\Delta}\lim_{T\rightarrow\infty}\frac{1}{ma_+T}\int_{-ma_+T/2}^{ma_+T/2}dx\,\sn^2\!\left(x,\frac{a_-}{a_+}\right).
}
The integrand is a periodic function over the real numbers, so we need to integrate over only one period in order to obtain the average
\al{\nonumber
C_{123}&=-\frac{16\pi}{3}\frac{m^2a_-^2}{\kappa}c_{\Delta}\frac{1}{2{\bf K}\big(\frac{a_-}{a_+}\big)}\int_{-{\bf K}\big(\frac{a_-}{a_+}\big)}^{{\bf K}\big(\frac{a_-}{a_+}\big)}dx\,\sn^2\!\left(x,\frac{a_-}{a_+}\right)\\
&=-\frac{16\pi}{3}\frac{m^2a_-^2}{\kappa}c_{\Delta}\frac{a_+^2}{a_-^2}\!\left(1-\frac{{\bf E}\big(\frac{a_-}{a_+}\big)}{{\bf K}\big(\frac{a_-}{a_+}\big)}\right).
}
We go back to real $m$, and finally get
\eq{
C_{123}=\frac{16\pi}{3}\frac{m^2a_+^2}{\kappa}c_{\Delta}\left(1-\frac{{\bf E}\big(\frac{a_-}{a_+}\big)}{{\bf K}\big(\frac{a_-}{a_+}\big)}\right).
\label{dilstrconstj}
}
As pointed out in \cite{Costa:2010,Bajnok:2014}, the structure constant should be proportional to the derivative of the string energy with respect to the square root of the 't Hooft coupling
\eq{
C_{123}=\frac{16\pi}{3}c_{\Delta}\frac{\partial E(J,I_{\theta},m,\sqrt{\lambda})}{\partial\sqrt{\lambda}}\,,
\label{derenergy}
}
where $I_{\theta}$ is the action variable corresponding to $\theta$. Differentiating the expression for $I_{\theta}$, obtained in \cite{Kruczenski:2004}, we are able to confirm the validity of \eqref{derenergy}.

Let us describe two particular cases of \eqref{dilstrconstj}. If we consider the case of large energy, namely large $\kappa=E/\sqrt{\lambda}$, we will get for the structure constant
\eq{
C_{123}=\frac{8\pi}{3}\kappa c_{\Delta}\left(1-\frac{8{\cal J}^2-m^2}{8\kappa^2}-\frac{4m^2{\cal J}^2-m^4}{16\kappa^4}+\ldots\right).
}
Another interesting case is when ${\cal J}\ll\kappa$. Then we get
\eq{
C_{123}\approx\frac{16\pi}{3}\kappa c_{\Delta}\!\left(1-\frac{{\bf E}\!\left(\frac{m}{\kappa}\right)}{{\bf K}\!\left(\frac{m}{\kappa}\right)}\right).
}

\subsection{Superconformal primary scalar as light operator}
The string state that corresponds to the chiral primary operator results from the trace of the graviton in the $S^5$ section of the geometry \cite{Kim:1985,Lee}. As detailed in \cite{Zarembo:2010,Berenstein:1998}, the bosonic part of the respective operator takes the form\footnote{We neglect derivative terms that will not influence our calculations since we have made the restriction ${\rm x}_1=-{\rm x}_2$.}
\al{\label{cpovertex}
V_L({\rm x}=0)&=V^{(\rm CPO)}_j(0)=\hat{c}_{\Delta}K_{\Delta}\,{\rm X}^j\big[(\partial x_m\bar{\partial}x^m-\partial z\bar{\partial}z)/z^2-
\partial X_k\bar{\partial}X_k\big]\,,\\
K_{\Delta}&\equiv\left(\frac{z}{z^2+x^mx_m}\right)^{\Delta}\!,\qquad{\rm X}\equiv X_1+iX_2=e^{i\varphi},\nonumber
}
where $\hat{c}_\Delta$ is again given by the normalization. The corresponding operator in the dual gauge theory is the BMN operator ${\rm tr}Z^j$ with dimension $\Delta=j$.

We can infer from \eqref{strconstst}, \eqref{norm3point} and \eqref{cpovertex} that
\al{
\label{cpostrconst}
C_{123}&=4c_{\Delta}\lim_{T\rightarrow\infty}\frac{1}{T}\int_{-T/2}^{T/2}d\tau_0\int^{\infty}_{-\infty}d\tau
\int^{\pi/2}_0d\sigma\,K_{\Delta}\,U\,e^{-(\Delta_2-\Delta_1)\kappa\tau_0}\\
U&={\rm X}^j\big[(\partial x_m\bar{\partial}x^m-\partial z\bar{\partial}z)/z^2-\partial X_k\bar{\partial}X_k\big],\qquad c_{\Delta}=2^{-\Delta}\hat{c}_{\Delta}\,,
}
where the constant $\hat{c}_\Delta$ of the superconformal scalar is \cite{Zarembo:2010,Berenstein:1998}
\eq{
\hat{c}_\Delta=\hat{c}_j=\frac{\sqrt{\lambda}}{8\pi N}(j+1)\sqrt{j}\,.
}
The expression for $U$, evaluated on the solution \eqref{sol}, leads to
\eq{
U=2\!\left(\frac{\kappa^2}{\cosh^2[\kappa(\tau-\tau_0)]}-\frac{{\cal J}^2}{\sin^2\theta}-m^2\cos^2\theta\right)\!
e^{ij\varphi_1},
}
so that \eqref{cpostrconst} gives
\eq{
C_{123}=4\pi c_{\Delta}\lim_{T\rightarrow\infty}\frac{1}{T}\!\int_{-T/2}^{T/2}\!\!\!\!d\tau_0\!
\int^{\infty}_{-\infty}\!\!\!\!d\tau\,\frac{e^{ij\varphi_1-j{\cal J}\tau_0}}{\cosh^j[\kappa(\tau-\tau_0)]}\!
\left(\frac{\kappa^2}{\cosh^2[\kappa(\tau-\tau_0)]}-\frac{{\cal J}^2}{\sin^2\theta}-m^2\cos^2\theta\!\right).
}
Analogously to the dilaton case the integrals cannot be calculated analytically, so we take ${\cal J}$ to be small and consider only the first term in the resulting series. We also change the variable $\tau_0$ to $\tau'=\tau_0-\tau$ and get
\al{\nonumber
C_{123}&=4\pi c_{\Delta}\int^{\infty}_{-\infty}\frac{d\tau'}{\cosh^j(\kappa\tau')} \lim_{T\rightarrow\infty}\frac{1}{T}\int_{-T/2}^{T/2}d\tau\left(\frac{\kappa^2}{\cosh^2(\kappa\tau')}-m^2\cos^2\theta\right)\\
&=4\pi\kappa^2c_{\Delta}\int^{\infty}_{-\infty}\frac{d\tau'}{\cosh^{j+2}(\kappa\tau')}-
4\pi m^2c_{\Delta}\int^{\infty}_{-\infty}\frac{d\tau'}{\cosh^j(\kappa\tau')} \lim_{T\rightarrow\infty}\frac{1}{T}\int_{-T/2}^{T/2}d\tau\,\cos^2\theta\nonumber\\
&=4\pi^{3/2}\kappa c_{\Delta}\frac{\Gamma[\frac{j}{2}]}{\Gamma[\frac{1+j}{2}]}
\left(\frac{{\bf E}\!\left(\frac{m}{\kappa}\right)}{{\bf K}\!\left(\frac{m}{\kappa}\right)}-\frac{1}{1+j}\right).
}

\sect{Conclusion}

The AdS/CFT correspondence has been through significant development in recent years. One of the active areas of research has been the holographic calculation of three-point functions at strong coupling. The correlation functions of three massive string states escape full comprehension so far~\cite{Klose:2011}, but we have uncovered almost all features of correlators containing two heavy and one light states in the semiclassical approximation~\cite{Zarembo:2010}--\cite{Hernandez:2010}.

In the present paper we calculated three-point correlation functions of two string and one supergravity states from string theory in $\axs$ at strong coupling, applying the approach of \cite{Roiban:2010} for computing correlators using the respective vertex operators. We examined the method, which had been correctly modified by the authors of \cite{Bajnok:2014}, for the occasion of a particular pulsating string solution, providing some limiting cases.

One of the possible future directions for exploration is the connection of our work to recent developments in the calculation of correlation functions with heavy states based on integrability methods in ${\cal N}=4$ SYM \cite{Escobedo:2010}.

\section*{Acknowledgments}
The authors would like to thank N. Bobev and H. Dimov for valuable discussions. This work was supported in part by the BNSF grant DFNI T02/6.

\end{document}